# Popular SQL Server Database Encryption Choices

Sourav Mukherjee
*Senior Database Administrator &
PhD student at University of the Cumberlands*
*Chicago, United States*
smukherjee3818@ucumberlands.edu

**Abstract**
*This article gives the overview of different database encryption choices in SQL Server. Which one works best in which situation. In today's world Data is more crucial than the expensive hardware cost. No one wants their personal data to be comprised. Same for business houses as well and they also do not want their data to be inappropriately handled to go out of the business. To help protect the public rights and safety, recently this year, the European Union had come up with strict rules and regulation of GDPR (General Data Protection Regulation).*

## I. INTRODUCTION

In today's world, data is more crucial than the expensive hardware cost. No one wants their personal data to be comprised. Same goes with business houses as well; they also do not want their data to be Inappropriately handled to go out of the business. To help protect the public rights and safety, the European Union had come up with strict rules and regulation of GDPR (General Data Protection Regulation) this year. Now, this regulation is confined to EU Economic Area and Territory. I'm sure other developed and developing countries will also bring their own territorial rules and guidelines sooner or later to strictly adhere to the data privacy rules in protecting confidential customer details. Anyway, this article is not related to the discussion about GDPR. However, the focus of this article is to talk about some of the excellent features developed by Microsoft to handle the data encryption, that is, how to implement more security features to the SQL Server software bundle (except the last feature).

The area of topic of discussion here is about the following list of four topics which are widely used and adopted by the organizations. By and large, every organization nowadays does have the Cybersecurity team. Further, many organizations have extended this arm to define their corporate security best practices. They do check out the ways the organization's valuable data assets can be protected and if they use Microsoft SQL Server to store data, then they use any, some, or all the below features. Now, Vormetric Transparent encryption is not developed by Microsoft. This is developed by Vormetric Inc, a San Jose based company that works on data security platform services. This product is tightly integrated to other encryption product keys, such as IBM's Guardium Data Encryption, Oracle's Transparent Data Encryption (TDE) and Microsoft SQL Server TDE, in addition to Vormetric keys.

Let's move on with the features of each of the four encryption models,

- Always Encrypted (AE)
- Transparent Data Encryption (TDE) & Cell Level Encryption (CLE)
- Dynamic Data Masking (DDE)
- Vormetric Transparent Encryption (VTE)

Now, when the entire IT world is concerned about data security, protection, vulnerability management, let's quickly check how Transparent Data Encryption (TDE) or Always Encrypted (AE) can help solve some of the problems.

The new Always Encrypted feature introduced in SQL 2016 allows the database administrators to encrypt sensitive data inside an application—without having to reveal the encryption keys to the SQL database or server.

Whereas, Transparent Data Encryption (TDE) and Cell Level Encryption (CLE) encrypt an entire database while at rest (at disk level), Always Encrypted encrypts at the column level but with several additional benefits (on the wire).

The client-side application is entirely uninformed of the implementation of TDE or the CLE and no such software is installed on the client-side system. All such encryption tasks are carried out by the SQL Server database alone.

The below diagram will show case about how a Transparent Encryption logic or algorithm functions.

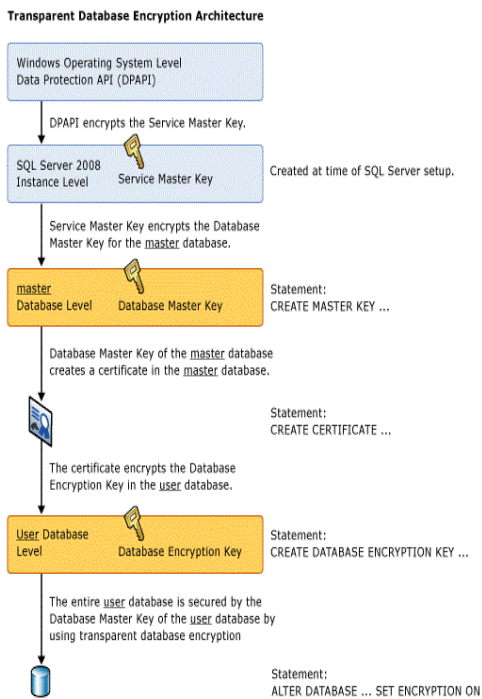

**Fig 1: Transparent Data Encryption Process [2]**

Basically, if you want your database backup to be protected then TDE works v well. If you implement TDE in the source server and if you want to restore database to another server then you need a master key, certificate to restore. Think about how you open your locker in the bank. One key is with you and the other key is with the ban professional to implement an extra layer on protection.

On the other hand, the Always Encrypted (AE) provides transparent encryption from the database to client applications. This AE feature is improved upon TDE by providing extra layer of encryption of sensitive data in memory and in transit, as well as at rest. The Always Encrypted-enabled driver actually performs the encryption and the decryption of the application. The owner of the information can then govern any potential leakage to database administrators by preserving the decryption keys so that administrators do not have incidental access to sensitive data. By contrast, the database administrator has access to the encryption keys with TDE by using the master key and the certificates.

The below diagram shows that how the Always Encrypted process typically works.

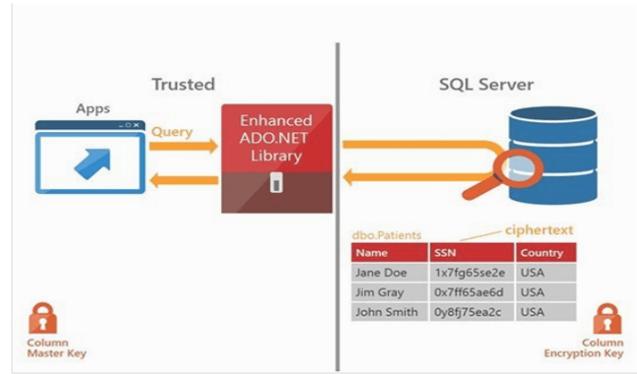

**Fig 2: Always Encrypted Process [1]**

Why I do prefer Always Encrypted in comparison with Transparent Data Encrypted is:

Always Encrypted can encrypt the data at the column level, rather than at the whole database level which is for Transparent Data Encrypted. So, you get a more granular choice to implementing the encryption. Is not that great?

- In Always Encrypted the data is encrypted both at rest and in the memory and where the decryption logic is defined at the Client side and it is done by Client Driver. This signifies that one can shield the data from harmful threats (by admins and by the man sitting in the middle to attack).
- Whereas Transparent Data Encryption (TDE) can only offer encryption at rest and that means not fully secured.

However, there are some limitations to the implementation of Always Encrypted.

- It can be just a single column which is so far for AE.
- To implement this feature, modification of the existing applications may be required.
- You can't involve tempdb to include part of AE
- Columns that are using one of the subsequent datatypes such as: xml, timestamp, rowversion, image, ntext, text, sql_variant, geography, hierarchyid, alias, geometry, user defined-types, etc.
- Some of the encrypted data types may require a "_bin2" collation type, which may require few DDL code changes.
- Your application will need to be compatible with .NET 4.6. If not, may not work.
- The application administrator will need to fully understand the encryption keys to ensure that they are protected—both from the

- Database administrators and other unintended audiences.
- The encryption keys will also need to be backed up for disaster recovery.
- Adding AE may increase your database size and CPU usage (especially for database writes) and adding encryption may also prevent any deduplication algorithms.
- Some of the SQL Server features such as replication are not currently supported in all editions. Instead you might have to upgrade to costlier Enterprise Edition.

One significant restriction of Always Encrypted is that it can only be applied to a limited subset of SQL operations. Many such SQL operations are complex and cannot be handled by Always Encrypted.

Also, note that–

- The Always Encrypted feature provides an entry-level additional level of security for sensitive data that may allow for reduction in administrator costs. Yet, the requirements tend to depend on new application development rather than tweaking the existing systems.
- Many complex SQL operations tasks may not work with Always Encrypted feature. I would only commend using Always Encrypted feature if the application design and architecture is fairly simple and straight forward.

For example, you may want to use Always Encrypted to send data from a SQL Server database which is hosted internally to a web-based SQL Server database and application. The data will be safe guarded in the transition and will be encrypted in the database. If your web application does basic SQL queries at the database layer, this method can work well.

**Dynamic Data Masking (DDM)**

Dynamic data masking is introduced in SQL Server 2016 edition and Azure SQL Database, and is configured by using basic Transact-SQL commands.

This feature is to restrict the revelation of sensitive data. It prevents the users by eliminating the access to the data to view it. This is a complementary security feature and need to be used with other security features such as audit, encryption or row level security.

Dynamic data masking (DDM) masks sensitive data on fly while protecting sensitive data to be viewed by non-privileged or designated users. It has few simple masking functions which are either inbuilt or you can customize based on your own need and through that you can prevent users to few unmasked data.

The beauty of this feature is that it does not require any coding effort from the application side or encrypting or applying any change to the real data stored in the disk.

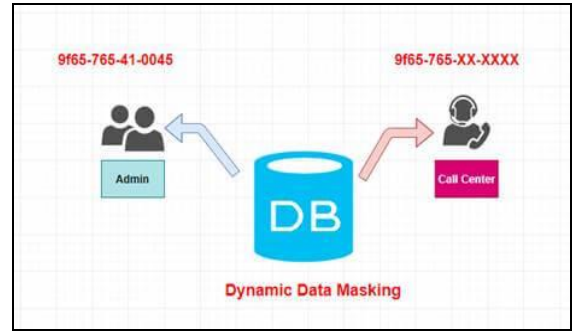

**Fig 3: Dynamic Data Masking Process**

One example of DDM which I can think of is that when calling the customer care of credit card company, they validate either last 4 or 6 digits of SSN number or any other sensitive customer data. If the application does support DDM, the call center agent may only ask few digits of the number rather asking for the entire sensitive number. This is a cool way of handling or protecting customer data.

There are four masking function in which are used to mask the data:

- Default
- Random

| Function | Description | Supported Type | Examples |
|---|---|---|---|
| Default | It completely masks the data. | Can work with all data types. | For string data types, use XXXX. |
| | | | For numeric data types use a zero value. |
| | | | For date and time data types use 01.01.1900 00:00:00.0000000 |
| | | | For binary data types use a single byte of ASCII value 0. |
| Custom String | Partially can mask the data. It exposes the first and last letters and adds a custom padding string in the middle. | Supports string data type | ALTER COLUMN [PhoneNo] ADD MASKED WITH (FUNCTION = 'partial(1,"XXXyyXXzzz",2)') |
| Email | It can mask the email type fields | Can work with all data types. | aXXX@XXXX.net |
| Random | It can mask the column values with random numbers. | It works with numeric data types | Examples includes numeric numbers such as phone number or anything related with numbers (example: SSN) |

- Custom String
- Email

**Code Snippet to be used to Implement Dynamic Data Masking using SQL Server T SQL (Edition SQL 2016 and onwards)**

--1 Create the Database
USE [master]
GO
CREATE DATABASE [DynamicDataMaskingDemo]
 CONTAINMENT = NONE
 ON  PRIMARY
( NAME = N'DDM', FILENAME = N'C:\Program Files\Microsoft SQL Server\MSSQL14.MSSQLSERVER\MSSQL\DATA\DDM.mdf' , SIZE = 8192KB , MAXSIZE = UNLIMITED, FILEGROWTH = 65536KB )
 LOG ON
( NAME = N'DDM_log', FILENAME = N'C:\Program Files\Microsoft SQL Server\MSSQL14.MSSQLSERVER\MSSQL\DATA\DDM_log.ldf' , SIZE = 8192KB , MAXSIZE = 2048GB , FILEGROWTH = 65536KB )
GO

--2 Create your table with proper functions
Use [DynamicDataMaskingDemo]
GO
CREATE TABLE [dbo].[EmployeeContacts]
(
        [ID] [int] IDENTITY(1,1) NOT NULL Primary key,
        [FName] [nvarchar](30) MASKED WITH (FUNCTION = 'default()') NOT NULL,
        [LName] [nvarchar](30) NOT NULL,
        [CreditCard] [varchar](20) MASKED WITH (FUNCTION = 'partial(2, "XX-XXXX-XXXX-XX", 2)') NULL,
        [SalaryUSD] [int] MASKED WITH (FUNCTION = 'default()') NULL,
        [OfficalEmail] [nvarchar](100) MASKED WITH (FUNCTION = 'email()') NULL,
        [PersonalEmail] [nvarchar](100) MASKED WITH (FUNCTION = 'email()') NULL,
        [ChangedDate] [datetime] MASKED WITH (FUNCTION = 'default()') NULL
)

--3 Insert Records to the table.
Insert Into [dbo].[EmployeeContacts] values ('Sourav', 'Mukherjee', '1234-3026-4508-4325', 10000,'smukherjee3818@ucumberlands.edu', 'smukherjee3818@ucumberlands.edu','2018-12-30 08:00:00')

--4 Grant permission to the user to the table.
CREATE USER [ReadOnlyTestUser] WITHOUT LOGIN;
GRANT SELECT ON [EmployeeContacts] TO [ReadOnlyTestUser];

--5 Check table records (with super user permission)
select * from [DynamicDataMaskingDemo].[dbo].[EmployeeContacts];

--6 Check the record using the user permission.
Execute as USer = 'ReadOnlyTestUser';
select * from [DynamicDataMaskingDemo].[dbo].[EmployeeContacts];
Revert
--This section will apply the masking function.

**Observation:**

- The data will be masked for the user ReadOnlyTestUser. Check out Fig #4 below about before masking vs after masking results.

**Conclusion:**

- This method allows developer to debug the Production environment without breaking the security or privacy for a user.

**Fig 4: Results of Dynamic Data Masking from SQL**

**Dynamic Data Masking also comes with its own limitations:**

A masking rule cannot be defined for the following column types:

- Columns with already defined encryption such as Always Encrypted

- Associated with FILESTREAM

- It cannot be configured on a computed column. However, if the computed column is dependent on a column with a MASK, then it will return the masked data.

- It can't be a part of Full Text Index

- It can't be performed on a column with any dependency. As a workaround, remove the dependency first and then add DDM and finally recreate the dependency. Let's say if the dependency is on a column in an index, then drop the index first then apply the mask and finally recreate the dependent index.

The final topic of discussion here is on Vormetric Transparent Encryption. Not sure how many of you got a chance to work using Vormetric but it offers many cool features.

- Vormetric is specialized for enterprise level encryption and for key management to protect databases which are in physical, virtual and Cloud Environments. This solution delivers comprehensive capabilities that enable your organization to address a broad range of security objectives. It also simplifies implementation and management for metric enables organizations to secure data across all their environments including physical virtual cloud and big data. The Data Security Platform contains encryption, the key management, key vault and toolkit products which are all managed via the Vormetric Data Security Manager also known as DSM. Vormetric's Transparent Encryption module is an agent which runs at the file system level on a server that encrypts data at rest. It is also used to access control, and for collecting security logs.

- It is often employed for compliance especially for protecting credit cards, medical records, personal information and intellectual property.

**Vormetric Product features**

Vormetric distinguishes itself from the competitors by providing transparent encryption, access controls that too at the granular levels and security intelligence. It also offers cloud platform support. It has the capability to encrypt data stored in all types of databases either it is structured or unstructured kind and at the file and folder levels which does not require any underlying changes to the databases or at the application level.

The Vormetric Transparent Encryption software has an agent which runs on servers or at the VM level to control access to files, folders and volumes. It finally reports the activities to DSM. The DSM is a physical or a virtual application which provides a Web-based user interface which can manage the complete platform. Agents can also apply the policies which are defined in the DSM to regulate the user actions, such as limiting which user can access the encryption key and for what purpose they can use it. The administrators can use the DSM to monitor the databases and systems. They can also view current status, or they can manage the encryption keys, or they can also govern access privileges.

Customers rely on Vormetric application encryption for several key reasons. By encrypting data in the application server, it is secured while in transit and in storage. Data is protected across its lifecycle including when it is backed up, migrated and archived. The solution insures that administrators and others with system and infrastructure access can't access sensitive data and application encryption can reduce the scope of PCI DSS compliance.

Very importantly if encrypted data is stolen attackers won't have any means to decrypt the data, so it won't be of any use or value even if a SQL injection attack is successful at extracting data from a compromised server only encrypted data will be returned to the hacker which won't be of any value and finally it protects from insider threats. Even your administrators won't be able to see sensitive.

## II. CONCLUSIONS

Every IT and security group are under growing pressure as they continue to have more sensitive data to protect, more threats to combat and more compliance mandates to address. One of the best ways to contend with these increasing security demands is to encrypt sensitive data. Recent research has revealed that how the adoption of encryption has increased as has the number of objectives it is being used to address the most common reasons for employing encryption.

In addition, encryption is being used to safeguard an organization's reputation.

Finally, security breaches have become a near certainty for most organizations now. Security managers started realizing that encryption can mitigate the damage that these attacks inflict to address. It is believed that more than thirty five percent of organizations have instituted a consistent encryption strategy. However, as the use of encryption has grown so has the number of encryption solutions that have been supported in fact more than seventy percent of organizations are working with five or more encryption technologies.

Companies historically have rolled out too many single purpose encryption tools and products that can only secure specific types of data systems or environments. As a result, to address their growing requirements, IT and security teams have had to procure, monitor and manage a collection of disjointed technologies which is complex, inefficient and expensive. Now organizations have a compelling alternative to these piecemeal approaches.

Under the hood, carefully choosing the right encryption policy and strategy is most important in keeping your organization safe and protecting the data and information secured.

## AUTHOR'S PROFILE

Sourav Mukherjee is a Senior Database Administrator and Data Architect based out of Chicago. He has more than 12 years of experience working with Microsoft SQL Server Database Platform. His work focusses in Microsoft SQL Server started with SQL Server 2000. Being a consultant architect, he has worked with different Chicago based clients. He has helped many companies in designing and maintaining their high availability solutions, developing and designing appropriate security models and providing query tuning guidelines to improve the overall SQL Server health, performance and simplifying the automation needs. He is passionate about SQL Server Database and the related community and contributing to articles in different SQL Server Public sites and Forums helping the community members. He holds a bachelor's degree in Computer Science & Engineering followed by a master's degree in Project Management. Currently pursuing Ph.D. In Information Technology from the University of the Cumberlands. His areas of research interest include RDBMS, distributed database, Cloud Security, AI and Machine Learning. He is an MCT (Microsoft Certified Trainer) since 2017 and holds other premier certifications such as MCP, MCTS, MCDBA, MCITP, TOGAF, Prince2, Certified Scrum Master and ITIL